\documentclass[review, 5p, twocolumn]{elsarticle}

\usepackage{lineno,hyperref}
\modulolinenumbers[5]

\journal{Journal of \LaTeX\ Templates}

\usepackage{url}
\usepackage[utf8]{inputenc}
\usepackage{tabularx}

\usepackage{orcidlink}
\usepackage{graphicx}
\usepackage{dcolumn}
\usepackage{bm}
\usepackage[utf8]{inputenc}
\usepackage[T1]{fontenc}
\usepackage{mathptmx}
\usepackage{etoolbox}
\usepackage{xcolor}
\usepackage{graphicx}
\usepackage{lineno,hyperref}
\usepackage{blindtext}
\usepackage{epsfig}
\usepackage{latexsym}
\usepackage{graphics}
\usepackage{epsfig}
\usepackage{amsmath,amssymb,amsfonts,theorem,color,bm}
\usepackage{multirow, multicol, boldline}
\usepackage{stfloats} 
\usepackage{float}
\usepackage{subfig}
\usepackage{soul}
\usepackage{caption}
\hypersetup{colorlinks,allcolors=black}
\usepackage[normalem]{ulem}

\newcommand\etal{{\it et al. }}

\begin{document}

\begin{frontmatter}

\title{Modeling heart flow dynamics using numerical simulations to identify the vortex ring: a practical guide.}

\newcommand{\orcidEL}{0000-0002-4514-6471}
\newcommand{\orcidAM}{0000-0002-3720-6710}
\newcommand{\orcidPQ}{0000-0003-4373-2079}
\newcommand{\orcidJGM}{0000-0002-7422-5320}
\newcommand{\orcidSLCM}{0000-0003-3605-7351}

\author[addressUPM]{E. Lazpita \orcidlink{\orcidEL}}
\cortext[mycorrespondingauthor]{Corresponding author}
\ead{e.lazpita@upm.es}

\author[addressUPV]{A. Mares \orcidlink{\orcidAM}}
\author[addressUPV]{P. Quintero \orcidlink{\orcidPQ}}
\author[addressUPM,addressCSC]{J. Garicano-Mena \orcidlink{\orcidJGM}}
\author[addressUPM,addressCSC]{S. Le Clainche \orcidlink{\orcidSLCM}}

\address[addressUPM]{ETSI Aeron\'autica y del Espacio - Universidad Polit\'ecnica de Madrid, 28040 Madrid, Spain}
\address[addressCSC]{Center for Computational Simulation (CCS), 28660 Boadilla del Monte, Spain}
\address[addressUPV]{ETSI Aeroespacial y Diseño Industrial - Universitat Politècnica de València, 46022 Valencia, Spain}

\begin{abstract}
In this study, we present a comprehensive numerical analysis of blood flow within human left ventricle models, with particular emphasis on optimizing simulation conditions to enhance the realism and computational efficiency of heart flow dynamics. The objective is to determine the most effective mesh configurations, flow conditions, and boundary settings necessary for accurately capturing the formation and behavior of the vortex ring—a pivotal element in ventricular flow dynamics. Utilizing a computational fluid mechanics approach, we review the influence of both idealized and patient-specific geometries on simulation outcomes. It is imperative to consider the necessity of dynamic wall motion and the precise calibration of inlet and outlet boundary conditions, which must be designed to mimic physiological conditions as accurately as possible. These factors are of paramount importance in achieving a balance between computational resource demands and the fidelity of the simulations, thereby providing valuable insights for future cardiovascular modeling efforts.
\end{abstract}


\end{frontmatter}


\section{\label{sec:Introduction} Introduction}

Despite considerable advances in cardiovascular medicine, cardiovascular disease (CVD) remains the leading cause of mortality worldwide \cite{Dorairaj2017}. This has motivated extensive research into innovative diagnostic and therapeutic strategies. The accurate detection and analysis of cardiac flow dynamics are fundamental to the advancement of cardiovascular medicine.
\\
Computational fluid dynamics (CFD) has emerged as a valuable tool for simulating cardiovascular flows. It offers a range of approaches and methodologies for modeling and analyzing the complex dynamics of blood flow within the heart. For instance, the open-source lifex-cfd solver demonstrates its utility in modeling blood flow under diverse physiological and pathological conditions. The study by Africa \etal \cite{Africa2024} underscores the solver's precision and dependability in simulations conducted through the incompressible Navier-Stokes equations. Another study presents a novel left ventricular simulator that employs computational modeling to test cardiovascular implants. This simulator has the potential to serve as a low-cost and efficient testing platform for preclinical evaluations \cite{Baturalp2024}. A comparative study of blood flow simulations using smoothed particle hydrodynamics (SPH) and finite volume method (FVM) methods has been conducted, and the findings indicate that while SPH provides detailed insights, it is more computationally demanding \cite{Topalovic2023}. Additionally, Zingaro \etal \cite{Zingaro2021} show the integration of electromechanics and fluid dynamics to achieve a realistic blood flow simulation in a left ventricle (LV) model. Finally, a study combining statistical shape modeling with CFD using cardiac computed tomography (CCT) data has been carried out to explore intracardiac flow features as early predictors of heart diseases \cite{Goubergrits2022}.

Despite its potential, CFD faces challenges in the accuracy and reliability when applied in cardiovascular research and eventually, in medical practice. While it is increasingly used to model heart flow dynamics, there is a lack of comprehensive guidelines for fine-tuning simulations to ensure optimal performance and credibility.
This study addresses these challenges by introducing a systematic approach to configure Star-CCM+ \cite{Starccm} simulations, a commercial solver used in the literature for heart flow dynamics \cite{Mohammad2013}. It places particular emphasis on accurately capturing the vortex ring, a critical flow feature in the left ventricle blood flow throughout the cardiac cycle.

The vortex ring is formed in the mitral valve of the left ventricle due to the jet entering through the orifice between the mitral valve leaflets. Initially, this vortex is planar, but as it enters an asymmetric cavity, it propagates and while doing so, experiences asymmetric growth and tilting during diastole. This is necessary to efficiently fill the ventricle chamber's elongated shape. This tilting is clockwise, with the left side remaining in the upper part due to the interaction of the ring with the ventricular walls, and the right side diving into the apex. This asymmetric development is crucial for efficient ventricular function, preventing stagnant areas and ensuring complete filling of the ventricle \cite{Schenkel2009}. An understanding of this relationship between ventricular shape, motion, and vortex evolution is essential for comprehending overall ventricular function \cite{Nguyen2015}.
\\

It has been demonstrated that variations in vortex ring formation and evolution can be linked to cardiovascular diseases. For instance, Mangual \etal \cite{Mangual2013} conducted a comparative evaluation of intraventricular fluid dynamics between healthy subjects and patients with dilated cardiomyopathy (DCM), identifying distinct differences in the vortical structures within the left ventricle. Similarly, Riva \etal \cite{Riva2024} used 3D magnetic resonance imaging (MRI) data to highlight differences in the characteristics of the vortex ring, such as circularity, orientation, and inclination, between healthy hearts and those with ischemic cardiomyopathy (ICM).
\\
The present study introduces a novel framework that improves the reliability and replicability of heart flow simulations. This development addresses a significant gap in the existing literature, wherein the available models frequently prove insufficient for accurately simulating complex cardiovascular dynamics with the desired level of replicability. This paper represents a significant advancement beyond a mere review of the literature. It effectively synthesizes the latest theoretical knowledge with practical applications. Consequently, our contribution represents a significant advancement over existing studies, offering unique insights and comprehension of the implementation selection, thereby facilitating more accurate and reproducible simulations. Furthermore, this article is enhanced by a practical guide, which can be found in Ref. \cite{ModelFLOWsCardiac}, which shows the optimal methodology for configuring CFD simulations, with a particular emphasis on the most intricate aspects of mesh generation and boundary condition settings. To the best of the authors' knowledge, there are currently no articles in the literature that provide researchers with a clear and comprehensive approach to addressing this issue. This will facilitate the exploration of new research avenues and the resolution of emerging problems.
\\
The article is organized as follows: Section \ref{sec:Conditions} details the methodology employed in tuning the simulations, presents the results for various conditions, and discusses their implications. Section \ref{sec:Validation} demonstrates the validation process using an alternative code implemented in Ansys Fluent \cite{Fluent}, as thoroughly explained in Ref. \cite{Lazpita2024ECCOMAS}. Finally, Section \ref{sec:Conclusions} summarizes the main conclusions of this work.

\section{\label{sec:Conditions} Simulation Setup}
In this section, we explore various conditions essential for optimizing our CFD simulations. The focus is on four key aspects: geometry of the model, flow modelling, mesh, and boundary conditions. Each of these factors plays a crucial role in the accuracy and efficiency when performing the simulations, particularly in their capacity to capture the vortex ring. This vortex is a pivotal indicator of efficient heart flow dynamics and requires precise simulation conditions to be depicted accurately. Our goal is to identify the simplest and most efficient methods for configuring these conditions, ensuring that they collectively contribute to a robust and reliable representation of the vortex ring in our simulations.

\subsection{\label{subsec:Geometry} Geometry}
\begin{figure}[h]
  \centering
  \includegraphics[trim=0 100 100 100, clip, width=0.45\textwidth]{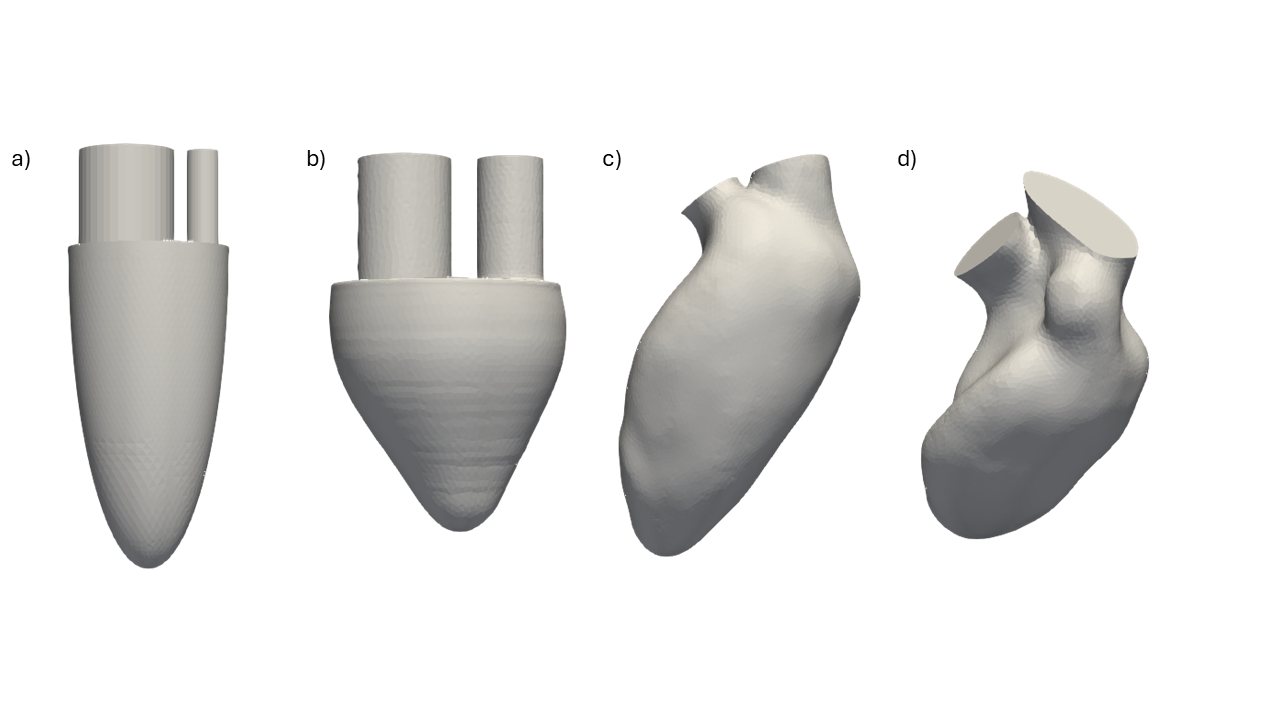}
  \caption{Left ventricle models, increasingly more realistic from left to right. Extracted from reference: a) Zheng \etal \cite{Zheng2012} and b) Vedula \etal \cite{Vedula2014}. Patient specific models extracted from medical data: c) Patient 1 and d) Patient 2.} \label{fig:geometries}
\end{figure}

Geometry plays a critical role in the accuracy of our results when performing the simulations. In our study, we utilized both idealized geometries and patient-specific geometries of the left heart. Idealized geometries were constructed based on Ref. \cite{Zheng2012, Vedula2014}, while patient-specific geometries were derived from medical imaging data. The first patient specific geometry was extracted from medical data available online as open source in \url{https://figshare.com/s/2a5de3a2b89a3fb87932} with its accompanying article \cite{CTscans}. The second patient data was obtained through a collaboration with the National Center of Cardiovascular interventions (CNIC) \cite{CNIC} and the project DigitHEART \cite{ModelFLOWsCardiac}. This different models are depicted in Fig. \ref{fig:geometries}.


Patient-specific geometries, when correctly represented, can provide more realistic and physiologically accurate results. However, these geometries also present several challenges. Obtaining these models is challenging due to the necessity of sophisticated medical imaging tools to create accurate CAD models of sufficient spatial resolution. Despite the availability of these tools, the segmentation and meshing process remains a significant challenge, both in terms of difficulty and automation. Additionally, the temporal resolution achieved during a heart cycle is often inadequate for the computational needs of our simulations, necessitating interpolation or other strategies to fill in data gaps. As a result, the person-cost of representing geometry and creating a mesh, along with the computational cost and complexity of the simulations, are increased.
\\
Conversely, idealized models, while potentially oversimplified, can be useful for capturing the most critical features of heart flow dynamics in a more controlled manner with reduced computational demand. These models are easier to validate due to their comparability and simplicity. Furthermore, idealized models facilitate the incorporation of various heart pathologies by directly modifying the model, avoiding the need to find a patient with the specific condition or to extensively alter a patient-derived model. This simplicity makes idealized geometries an attractive option for simulations where control and cost are significant considerations. Therefore, for the sake of brevity, through the following discussion we will mainly focus on the idealized geometry extracted from Zheng \etal \cite{Zheng2012} selected from references for its simplicity when configuring our simulation.
\\
We represent this model of the left ventricular cavity as a semi-ellipsoid with two cylindrical tubes attached in the top plane, which model the mitral and aortic valves. To accurately model the blood flow within a human heart, it is essential to select the appropriate geometrical parameters, ensuring a consistent end-systolic volume (ESV) for the LV model, which represents the minimum volume of the heart during a cycle. 
These include the semiaxis lengths ($a = 2 cm$ and $b = 8cm$), the tube diameters ($D = 2.4 cm$ and $d = 0.8 cm$ for the inlet and the outlet, respectively), the height of the tubes ($H$), and finally, the distance between the centers of the tubes and the axis of the semi-ellipsoid ($C = 0.55 cm$ and $c = 1.35 cm$ for the inlet and outlet, respectively). Figure \ref{fig:geometry} details further the geometry and the location of each geometrical parameter necessary to contruct the model. With these dimensions, the volume in the ESV state for this model, without taking into account the tubes, can be calculated using the volume equation for a semi-ellipsoid with a circular base:
\begin{equation}
    ESV = 2/3 \cdot \pi \cdot a^2 \cdot b = 67 \mathrm{ ml}
    \label{eq:ESV}
\end{equation}

The values of the dimensions, and consequently, the volume, are selected to match physiological values. The evolution of the volume of this geometry from ESV to end diastolic volume (EDV) and back is performed moving the outer wall of the semi-ellipsoid, which is further explained in Section \ref{subsubsec:Wall}.

\begin{figure}[h]
  \centering
  \includegraphics[trim=0 0 200 0, clip, width=0.45\textwidth]{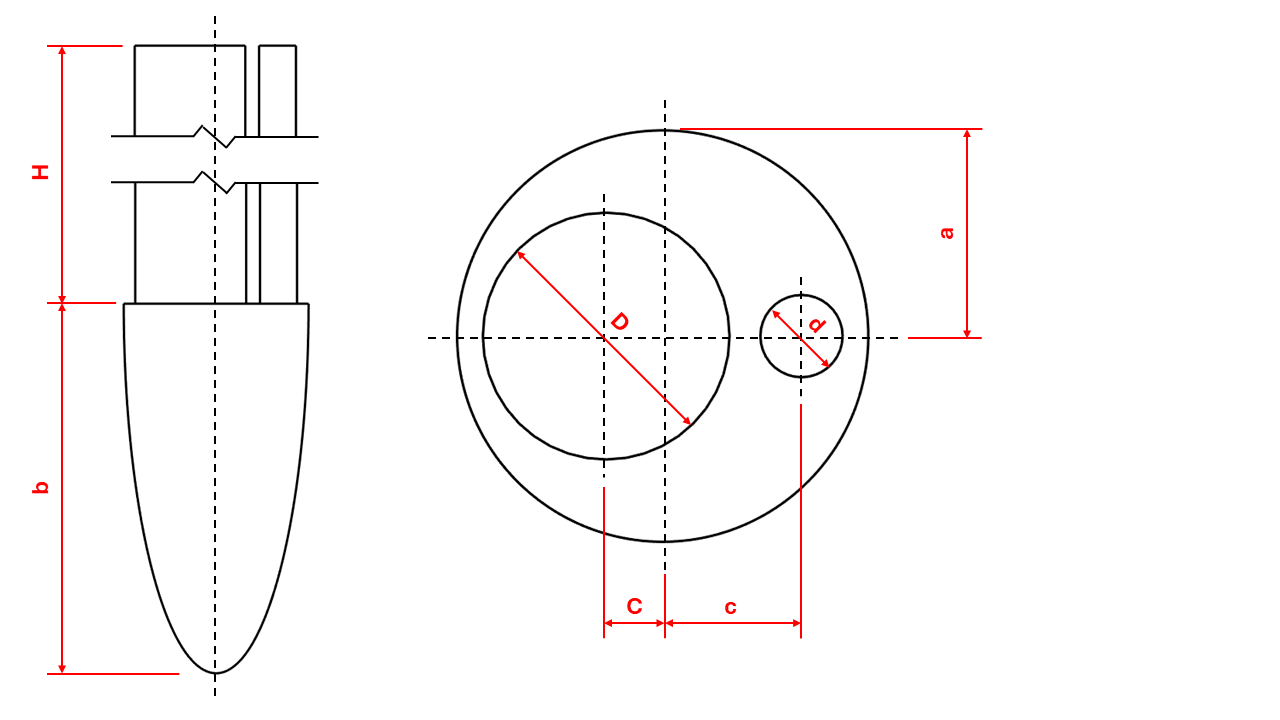}
  \caption{Ideal geometry of the LV model extracted from Zheng \cite{Zheng2012}.} \label{fig:geometry}
\end{figure}

\subsection{\label{subsec:FlowConditions} Flow Modelling}
Selecting appropriate fluid conditions is of paramount importance for the accurate simulation setup. In our study, we model blood as an incompressible Newtonian fluid with a constant density ($\rho = 1060$ $kg/m^3$) and dynamic viscosity ($\mu = 0.004$ $Pa \cdot s$). The Reynolds number can be inferred from these values by taking the inlet tube diameter, $D$, as the characteristic length and the maximum inlet velocity of the entire cardiac cycle ($V_{max} = 0.86 m/s$) as the characteristic velocity. This will be discussed in greater detail in Section \ref{subsubsec:Inlet}. Consequently, the Reynolds number can be defined as:
\begin{equation}
    \mathrm{Re} = \frac{\rho V_{max} D}{\mu} \simeq 5500
\end{equation}
\\
Another typical non-dimensional parameter employed in in cardiovascular simulations is the Womersley number, $\alpha$, which quantifies the ratio of transient inertial forces to viscous forces and characterizes the pulsatile nature of blood flow. The Womersley number is defined as: 
\begin{equation}
    \alpha = D \left( \frac{\omega \rho}{\mu} \right)^{1/2} \simeq 31
\end{equation}
where $\omega$ is the angular frequency of the heart cycle which in our case is $2\pi$. This value is greater than that observed in Ref. \cite{Simpson2023}, since the characteristic length is approximately ten times larger.
\\
Given that the blood flow within the LV chambers is characterized by a regime of transition to turbulence, it is essential to include turbulence models in a detailed simulation that captures a wider range of scales and phenomena. Consequently, certain studies, such as those conducted by those by Bucelli \etal \cite{Bucelli2023} and Korte \etal \cite{Korte2023}, employ turbulence models (Variational Multiscale - Large Eddy Simulation and Shear Stress Transport $k-\omega$ model, respectively) to capture the entirety of the features that manifest throughout the cardiac cycle, thereby facilitating the generation of highly precise simulations.
\\
However, our objective is to capture the vortex ring, a phenomenon that emerges in early diastole with low velocity. In light of the findings presented in the literature, namely those of He \etal \cite{He2022} and Tagliabue \etal \cite{Tagliabue2017}, it is possible to accurately and efficiently capture the formation and evolution of the vortex ring by assuming that the flow behaves as laminar and not considering turbulence in the simulations.

\subsection{\label{subsec:Mesh} Mesh}
The process of meshing plays a pivotal role in our simulations, particularly in light of the intricate dynamics of fluid flow within a cardiac chamber. In these confined spaces, we encounter a multitude of small-scale phenomena that necessitate the use of a highly refined mesh to accurately capture the relevant flow dynamics.


\begin{figure}[h]
  \centering
  \includegraphics[trim=0 0 350 0, clip, width=0.45\textwidth]{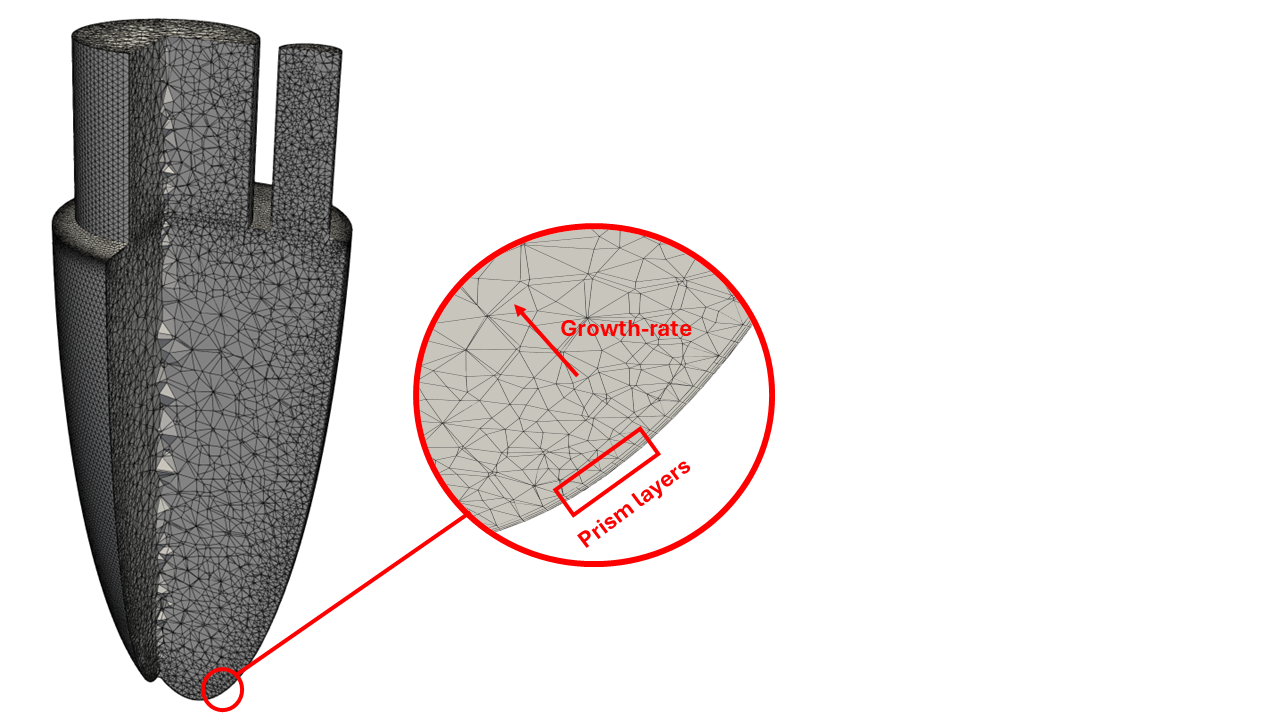}
  \caption{Illustration of the mesh for the properties detailed in Tab. \ref{tab:mesh} and a base size of $7.5 \cdot 10^{-4} m$.} \label{fig:geometry}
\end{figure}


To create our meshes, we changed some properties in the Star-CCM+ meshing tool, which are described in Tab. \ref{tab:mesh}. First of all, we need to declare a base size which will be the default size of an element in our mesh. This size will vary over the volume depending on other properties such as the growth rate. In our study, we selected a $20 \cdot 10^{-4} m$ size to start with and then, in order to achieve mesh convergence, the base element size was decreased to obtain finer meshes. We then implement a mesh refinement approach that includes two primary strategies: applying a tetrahedral mesh to most of the domain at a given volumetric growth rate, ensuring that the mesh becomes increasingly refined as it approaches the walls, and incorporating four to six prismatic layers attached to the walls. These prism layers are particularly effective at adapting the mesh to the contours and movements of the walls, thus increasing the accuracy of our simulations. An auto-generative iterative process was employed in order to obtain a conformed and high-quality mesh.
\\
It should be noted that the meshing process is not typically depicted in the literature, as it is a complex procedure when attempting to reproduce the findings of another study. This is because it necessitates a profound understanding and expertise with the solver. In Ref. \cite{ModelFLOWsCardiac}, we present the methodology employed to tune our meshes.
\\
This meticulous approach to mesh configuration is essential for capturing relevant phenomena at the wall boundaries, and also paves the way for the integration of wall motion and mesh transformation techniques discussed in the next section.


\begin{table}[h]
    \centering
    {\normalsize
    \begin{tabular}{|l l|}
            \hline
            Property                 & Value                             \\
            \hline \hline
            Type                     & Hybrid Tetrahedral/Prism         \\
            Base size [m]           & [20, 10, 7.5, 5] $ \cdot 10^{-4}$ \\
            Volume growth rate       & 1.3                               \\
            Nº prism layers          & 5                                 \\
            Prism layers growth rate & 1.2                               \\
            \hline
        \end{tabular}
    }
    \caption{Mesh properties in the Star-CCM+ meshing tool. The base size reflects the standard size of the elements in our mesh, the volume growth rate specifies the rate at which the element size grows as we separate from the outer surface, the number of prism layers selects how many prism layers the user wants to add around the surface, and again the growth rate specifies the rate at which they grow.} \label{tab:mesh}
\end{table}


\subsection{\label{subsec:BoundaryConditions} Boundary Conditions}
In our study, which focuses on an idealized LV model, we consider three primary boundaries: the ventricular wall, which can include its displacement; the inlet, which simulates the blood flow entering through the mitral valve; and the outlet, which represents the flow exiting through the aortic valve. Each boundary plays a pivotal role in accurately representing the physiological processes of the heart, thus necessitating precise definition to ensure the realism and success of our simulations.

\subsubsection{\label{subsubsec:Wall} Wall}
\begin{figure}[h]
  \centering
  \includegraphics[trim=0 20 150 0, clip, width=0.45\textwidth]{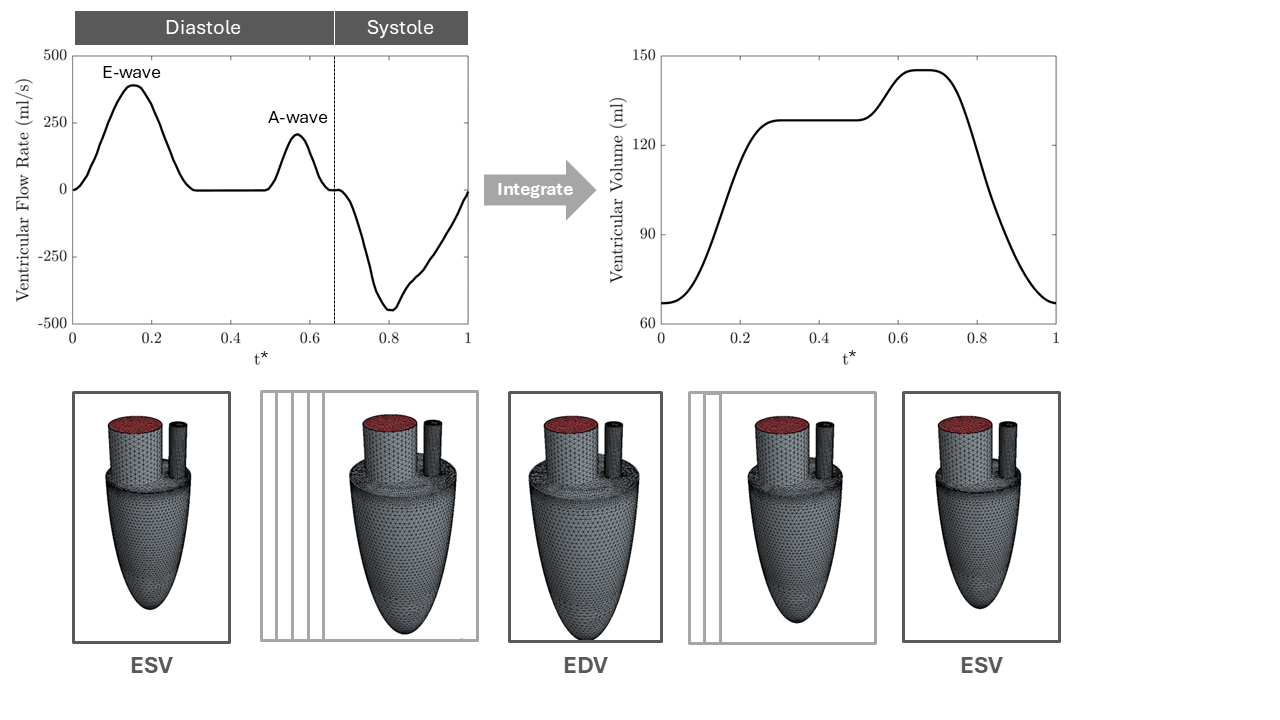}
  \caption{(Top) A detailed plot of the volumetric flow rate of the LV model during a heart cycle, extracted from Ref. \cite{Zheng2012} and the volume variation graph of the same model obtained via integration in Matlab are provided herewith. (Bottom) Representation of the various mesh states from end-systolic volume (ESV) at time point zero to end-diastolic volume (EDV) at time point 0.67 and back to ESV at time point one.} \label{fig:mesh_movement}
\end{figure}

In the study of left ventricle hemodynamics, the selection of wall boundary conditions is a matter of careful consideration. The decision to employ fixed walls or periodically moving walls, which more closely reflect the dynamic nature of the heart, is of critical importance, yet it introduces varying degrees of complexity to the implementation and analysis. At first, the walls were considered fixed throughout the cardiac cycle with the objective of approximating inflow and outflow rates to match idealized or physiological data. However, our initial experiments demonstrated that fixed walls were unable to accurately model the flow dynamics within the ventricle; key features such as the vortex ring or recirculation to the apex were not captured. This highlights the limitations of the fixed wall configuration in representing the intricate flow patterns within the ventricular chamber.
\\
In light of the critical importance of capturing accurate dynamics, we turned to an implementation of moving walls. However, this posed significant challenges. Figure \ref{fig:mesh_movement} illustrates a schematic approach of the methodology used to obtain the wall movement, showing the volume flow rate and volume change graphs in the top row and some time instants of the evolution of the mesh during a cardiac cycle.
\\
In this study, we employed flow rate data extracted from Zheng \etal \cite{Zheng2012}, which represents the volume of blood entering the left ventricular chamber per unit of time during a heart cycle. In our approach, the heart rate was assumed to be 60 beats per minute, so the heart period was $T = 1s$. From now on, time will be normalized to the period as $t^* = t/T$ for generalization purposes. Given the assumption of incompressibility for the blood flow, integration of the flow rate data allows the determination of the volume change of the LV model for each time instant. To this end, the raw flow rate data extracted from the reference was first interpolated into a temporally equispaced grid using the "\texttt{interp1}" function with the "\texttt{spline}" method. Subsequently, the integral was calculated using the "\texttt{integrate}" function, which employs a numerical integration method. The aforementioned process and methodologies were implemented in Matlab \cite{Matlab} and are illustrated in the upper panel of Fig. \ref{fig:mesh_movement}.
\\
Once the evolution of ventricular volume $V(t^*)$ has been calculated, the long and short semiaxis values for each time instant can be calculated using the semi-ellipsoid volume equation and maintaining the 4 to 1 ratio ($k=4$) between $b(t^*)$ and $a(t^*)$:
\begin{equation}
    b(t^*) = \left[ \frac{3}{2\pi} \cdot k^2 \cdot V(t^*) \right]^{1/3}
\end{equation}
\begin{equation}
    a(t^*) = k \cdot b(t^*)
\end{equation}
\\
The geometry was initially constructed as the minimum volume of the ventricle, the end-systolic volume (ESV), at time $t^*=0$. It then undergoes a uniform expansion following the previous equations to attain the maximum volume, the end-diastolic volume (EDV), at time $t^*=0.67$, before reverting to the ESV at time $t^*=1$. This is illustrated in the bottom of Fig. \ref{fig:mesh_movement}.
\\
By obtaining the volume change and defining the geometry for each time instant, we have gathered the required data for the implementation of the wall movement. The data will be utilized by each solver or tool in a manner contingent upon the implementation of the code. For example, Ansys Fluent necessitates the introduction of each time instant of the mesh with User Defined Functions. In our study employing Star-CCM+, it is initially essential to normalize the volume variation between 0 (ESV) and 1 (EDV) to employ the data in the "\texttt{morphing}" tool as incremental displacements. Subsequently, utilizing the "\texttt{diff}" function of Matlab, the displacement of the mesh was calculated in order to align with these changes, effectively morphing it to adapt to the dynamic volume alterations.


The comprehensive manipulation of this data has facilitated the implementation of the wall movement in Star-CCM+. The initial step is to construct or import the model in its minimum volume instant. Subsequently, the model must be defined in its maximum volume state and the total displacement from the minimum to the maximum volumes must be extracted from the mesh states obtained earlier. Finally, the incremental displacements between each time step must be introduced as a morphing wall boundary condition to guide the morphing process for the volume changes to match the flow rate graph.
\\
This implementation enabled the simulation of ventricular wall movement in a manner that approximates the physiological reality. The simulation successfully captured key features, such as the vortex ring and other essential characteristics of cardiac flow within the ventricle. 

\subsubsection{\label{subsubsec:Inlet} Inlet}
The selection of appropriate inlet boundary conditions is of critical importance, particularly in the context of modeling the flow entering the idealized LV model. Our objective is to accurately reproduce the dynamics of this flow, taking into account a number of factors, including the length of the inlet tube, the type of boundary, and the temporal and spatial distribution of the incoming flow.

\begin{figure}[h]
  \centering
  \includegraphics[trim=40 50 190 0, clip, width=0.45\textwidth]{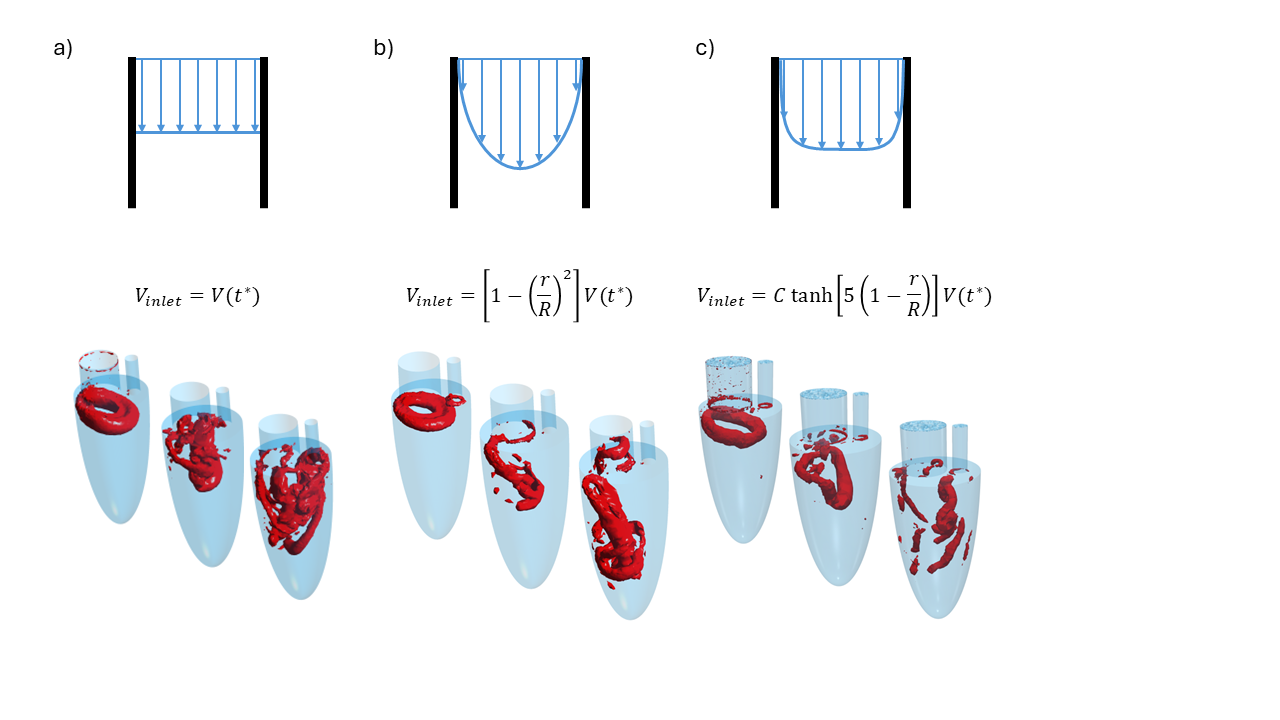}
  \caption{Each column represents a different inlet spatial profile: a) Plug; b) Parabolic; c) Hyperbolic tangent. (Top) Schematic representation of the inlet profile in a 2-d section. (Center) Equation for the velocity inlet spatial profile. (Bottom) Q-criterion isosurfaces for $t^* = 0.16$, $0.26$, $0.40$ from left to right, depicting the evolution of the vortex ring.} \label{fig:inlet_profiles}
\end{figure}

Our initial experiments explored different boundary condition types, such as pressure inlet, velocity inlet, and mass flow inlet. A pressure-only boundary condition was initially employed, which permitted the flow to proceed inside the ventricle due to the pressure gradients created by ventricular expansion. In our experiments with this boundary condition, we observed suboptimal performance in capturing the relevant flow patterns and encountered difficulties in modeling the mechanical actions of the mitral valve (opening and closing), which are essential for accurate modeling of intracardiac flows.
\\
In contrast, the utilization of either a velocity or mass flow inlet proved more effective in our study in accurately capturing the valve closure and overall flow dynamics. However, the performance of these approaches is contingent upon the precise definition of the flow profile, both temporally and spatially. In terms of temporal characteristics, the flow profile must closely resemble the physiological data, exhibiting a large peak corresponding to the E-wave, followed by a smaller peak for the A-wave, and ceasing during valve closure. This information was extracted from the flow rate graph in Fig. \ref{fig:mesh_movement} and modified for zero flow/velocity in systole.
\\
In regard to the spatial distribution of the profile, the most common initial choice is the plug profile, which employs a uniform distribution. However, due to the presence of a wall-bounded tube, a discontinuity in the flow may occur in the immediate vicinity of the wall, where both zero velocity (no-slip wall boundary condition) and a velocity profile are imposed. Accordingly, we have also included profiles that may mitigate this discontinuity, such as a parabolic profile or a midway approach as the hyperbolic tangent, wherein the velocity in the wall is set to zero and then grows to a value approaching that of the plug profile for the tube center. In all cases, the flow rate was maintained constant, with the necessary corrections being introduced in the equations. Consequently, our study will encompass three distinct spatial profiles, as illustrated in Fig. \ref{fig:inlet_profiles}:

\begin{equation}
    V_{plug} = V(t^*)
\end{equation}
\begin{equation}
    V_{parabolic} = \left[ 1 - \left( \frac{r}{R} \right)^2 \right] V(t^*)
\end{equation}
\begin{equation}
    V_{tanh} = C \cdot tanh\left[ 5 \left( 1 -  \frac{r}{R} \right) \right] V(t^*)
\end{equation}
where $C$ is the constant used to match the flow rates.
\\
The hyperbolic tangent distribution proved to be the most effective since it improved convergence on coarser meshes, reducing computational cost, while capturing the main dynamics of the vortex ring, namely the formation of the circular ring in early diastole, a clockwise tilt due to the proximity of the wall on one side of the ring and the reduction of the inflow velocity, and finally the instability growth and final dissipation of the ring at the end of diastole. The plug profile also captured the main dynamics just mentioned, but proved to be more computationally demanding, requiring finer meshes to achieve convergence. Finally, the parabolic profile proved to perturb the incoming flow excessively, causing the ring to reach too deep, thus slowing down the tilt, enlarging the structure and making the growth and dissipation of the instability slower and later in the cycle, thus affecting the overall flow dynamics. In Figure \ref{fig:inlet_profiles} we can see these results illustrated, showing the schematic of the different inlet profiles just at the beginning of the inlet tube in a 2-d section at the top, the equation in the middle and three different time instants ($t^* = 0.16$, $0.26$ and $0.40$ from left to right) of the Q-criterion isosurfaces inside the left ventricle model.


Furthermore, the impact of tube length on flow quality was assessed by testing lengths of $L = 1D$, $2D$, and $5D$, where $D$ represents the inlet tube diameter. The findings indicate that a tube length of at least $L = 2D$ is necessary to ensure a correct flow profile before it enters the ventricle.

In conclusion, the optimal configuration for our simulations incorporates a velocity or mass flow inlet with a temporal profile based on reference data, employs a hyperbolic tangent spatial distribution for idealized geometries (or a plug profile for patient-specific cases), and uses an inlet tube length of at least $2D$. This combination has proven to be the most efficient in capturing the vortical features within the LV model.

\subsubsection{\label{subsubsec:Outlet} Outlet}

In order to simplify our model, we matched the tube length at the outlet to that of the inlet. This decision was based on the understanding that the outlet length has minimal impact on the results, as it is mainly associated with the outflow and does not directly affect the formation of the vortex ring.
\\
The primary consideration at the outlet is the selection between a free pressure outlet and a mass flow outlet with a controlled time profile. A free pressure outlet would necessitate the implementation of a mechanism to simulate the periodic opening and closing of the aortic valve. One potential approach would be to temporarily convert the boundary to a wall during diastole (when the heart relaxes) and then switch back to a pressure outlet during systole (when the heart contracts). However, our experimental findings have indicated that this methodology can potentially introduce noise and instability immediately following these transitions. Nevertheless, these disturbances tend to stabilize after a few time steps.
\\
An alternative option was considered, namely the utilization of a mass flow outlet. This approach facilitates the modeling of aortic valve closure by enabling the flow to be set to zero during diastole without modifying the boundary conditions. Furthermore, the mass flow outlet offers enhanced control over the flow during systole, enabling precise matching of the flow rate to the volume changes in the left ventricle (LV), which helps prevent numerical instabilities.
\\
In conclusion, the findings of our study indicate that a mass flow outlet represents the optimal choice for controlling outflow dynamics in LV simulations. This is due to the fact that it permits straightforward programming of zero outflow during diastole and facilitates more precise control of the flow rate during systole, thereby reducing the likelihood of errors.

\section{\label{sec:Validation} Validation}

\begin{table*}[h]
    \centering
    {\normalsize
    \begin{tabular}{|l l l|}
            \hline
            \texttt{Commercial} solver       & Fluent                                     & Star-CCM+                            \\
            \hline \hline
            Density                 & $\rho = 1060$ $kg/m^3$ $= cte$             & $\rho = 1060$ $kg/m^3$ $= cte$       \\
            Viscosity               & $\mu = 0.004$ $Pa\cdot s$ $= cte$         & $\mu = 0.004$ $Pa\cdot s$ $= cte$   \\
            Flow regime             & Laminar                                    & Laminar                              \\
            Inlet                   & Velocity inlet (diastole), wall (systole)  & Velocity inlet, temporal profile     \\
            Outlet                  & Wall (diastole), pressure outlet (systole) & Mass flow outlet, temporal profile   \\
            Wall movement           & UDF to introduce each time instant mesh    & Morphing from ESV to EDV             \\
            Temporal discretization & 2nd order                                  & 2nd order                            \\
            Spatial discretization  & 2nd order                                  & 2nd order                            \\
            \hline
        \end{tabular}
    }
    \caption{Most relevant simulation setup for the different solvers.} \label{tab:simulation_setup}
\end{table*}

\begin{figure}[h]
  \centering
  \includegraphics[trim=0 0 0 0, clip, width=0.45\textwidth]{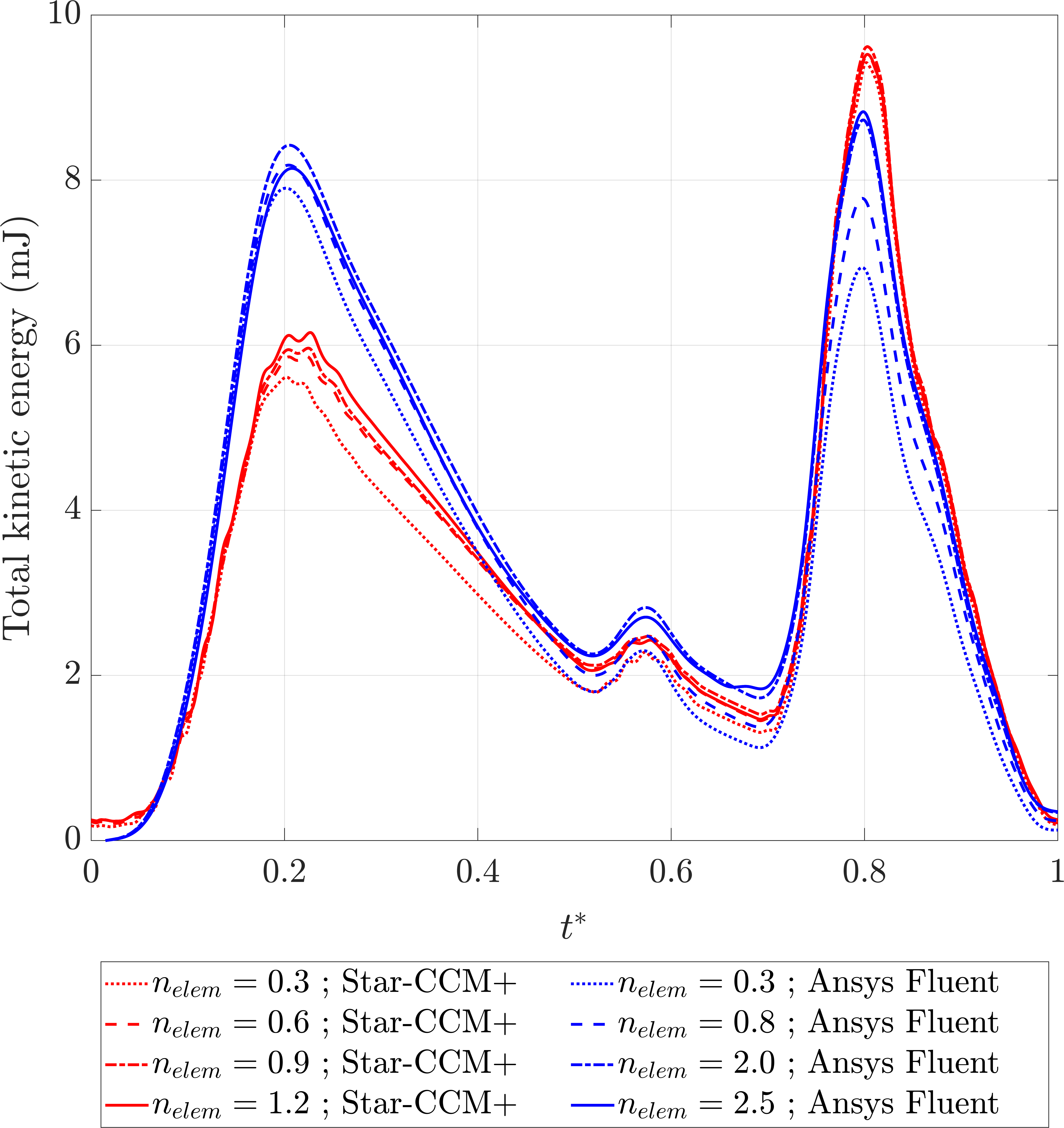}
  \caption{Evolution of the TKE for a heart cycle for different mesh sizes.} \label{fig:TKE}
\end{figure}

\begin{figure}[h]
  \centering
  \includegraphics[trim=0 0 250 0, clip, width=0.45\textwidth]{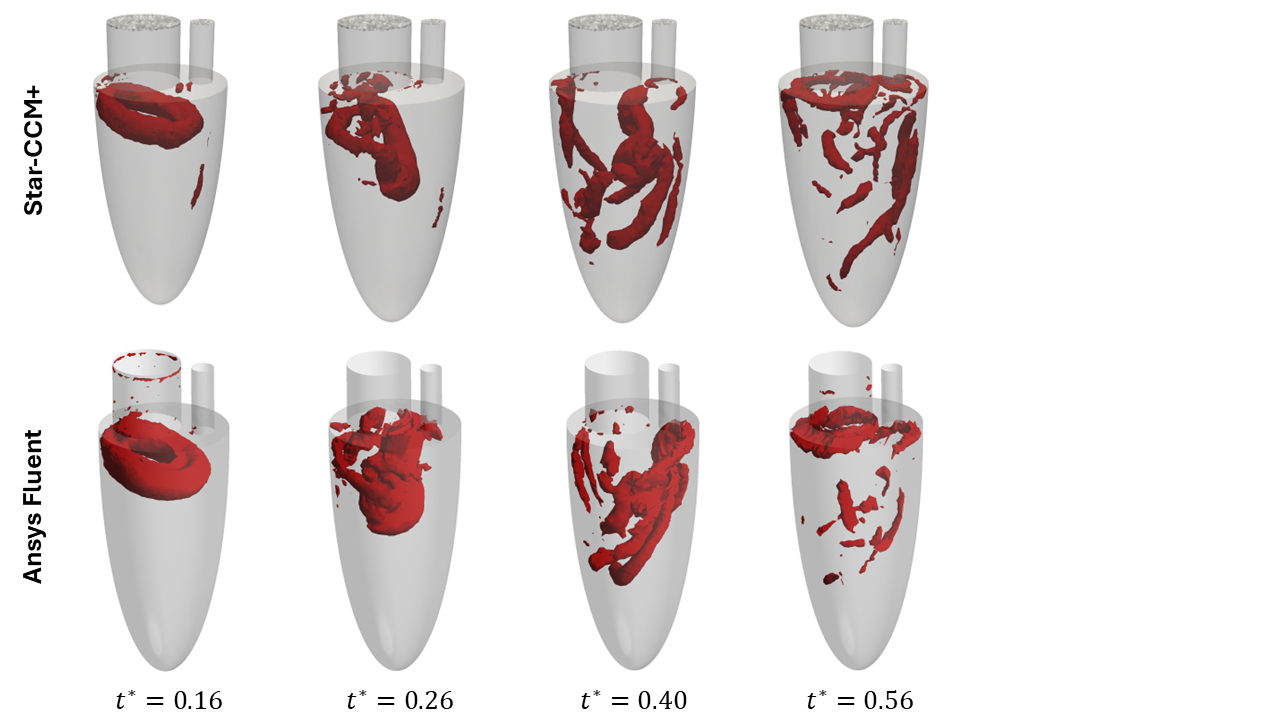}
  \caption{Isocontours of the Q-Criterion inside the ventricular cavity that show the evolution of the vortex ring for both solvers during diastole.} \label{fig:Qcriterion}
\end{figure}

To ensure the accuracy and efficiency of our simulation results, in Ref. \cite{Lazpita2024ECCOMAS} we showed the validation of our findings using two different CFD solvers implementations: Ansys Fluent \cite{Fluent} and Star-CCM+ \cite{Starccm}. For the sake of clarity, we provide a summary of the validation and implementation differences. Moreover, this article also introduces the capability of the present setup to properly follow the temporal evolution of the vortex ring.
\\
Table \ref{tab:simulation_setup} provides a detailed overview of the various options available for each solver's simulation setup. The primary distinctions in boundary condition implementation are as follows: the Fluent solver employs a wall boundary condition at both the inlet and outlet during diastole and systole, respectively. Additionally, a pressure outlet is employed when the aortic valve is opened. The wall movement is coded differently due to the inherent limitations of each solver. For instance, Fluent requires the introduction of a mesh at each time instant, whereas Star-CCM+ utilizes morphing to transition from the minimum to the maximum volume and back.
\\
We evaluated the convergence of our simulations using three different mesh sizes in both Fluent and Star-CCM+. These included coarse (0.3 million and 0.6 million cells), medium (0.8 million and 0.9 million cells), and fine (2.0 million and 1.2 million cells) meshes, respectively. Our primary metric for validation was the total kinetic energy (TKE) within the ventricular cavity, excluding the inlet and outlet tubes, over one cardiac cycle.
\\
This metric is shown in Fig. \ref{fig:TKE} for both commercial solvers. The results show a consistent pattern with small variations in peak values, consistent with previously published data \cite{Vedula2014}. As we refined the mesh, we observed that the TKE distributions for both Fluent and Star-CCM+ converged to consistent profiles. This convergence can also be seen in Tab. \ref{tab:mesh_convergence} where we display the relative error between the current mesh size and the finest. For medium meshes the error is below 2\% for both solvers, indicating that our simulations are reliable and that finer mesh resolutions improve the accuracy of the results.
\\
Furthermore, our research concentrated on the formation and evolution of the vortex ring, a topic that has been extensively examined in the existing literature (for further details, please see Ref. \cite{Le2012}). Figure \ref{fig:Qcriterion} illustrates this process, showing the results for Star-CCM+ on the top row and the same isosurfaces for the Ansys Fluent solver on the bottom row for four different time instants during diastole. Firstly, in the initial phase of diastole (at time point $t^*=0.16$), the vortex ring is formed at the exit of the inlet tube, which represents the mitral valve, when it opens during the E-wave. In the initial phase, the vortex ring assumes a circular configuration. As it progresses through the ventricle, at time $t^*=0.26$, the ring exhibits a tilting motion, influenced by the proximity of the wall of one of its sides and the accompanying reduction in inflow velocity. At time $t^*=0.40$, interactions between the vortex ring and the ventricular walls result in the formation of secondary vortex tubes, which generate complex instabilities. These instabilities tend to grow, and by the time of the latter diastole ($t^*=0.56$), they have caused the vortex ring to dissipate. In its place, a second, smaller vortex ring has appeared in the mitral valve due to the A-wave. It is noteworthy that the results presented in Fig. \ref{fig:Qcriterion} obtained with both solvers are in close agreement with each other and with the documented results in Ref. \cite{Zheng2012} Fig. 3.
\\
By validating our simulations using two different CFD solver implementations, we have demonstrated the robustness and accuracy of our approach in capturing critical hemodynamic features, such as the vortex ring, which strengthens the reliability of our findings. Furthermore, we have provided a highly detailed explanation of all critical aspects of the CFD simulation setup compared to other studies in the literature. This includes geometry construction, flow modeling, wall motion, and implementation of inlet and outlet conditions. Our decisions and results are thoroughly supported by references. In addition, this article is complemented by an online tutorial available on our website \cite{ModelFLOWsCardiac}, which facilitates the reproduction of consistent results.


\begin{table*}[h]
    \centering
    {\normalsize
    \begin{tabular}{V{2.5} l | c | c | c | c V{2.5}}
            \hlineB{2.5}
            \multirow{2}{*}{Mesh size}  & \multicolumn{2}{c |}{Star-CCM+} & \multicolumn{2}{c V{2.5}}{Ansys Fluent} \\
            \cline{2-5}
                                        & Nº elements    & Relative error (\%)           & Nº elements    & Relative error (\%)            \\
            \hlineB{2.5}
            Very coarse                 & 0.3         & 7.82                     & 0.3         & 13.62                     \\
            Coarse                      & 0.6         & 2.06                     & 0.8         & 5.59                      \\
            Medium                      & 0.9         & 0.42                     & 2.0         & 1.72                      \\
            Fine                        & 1.2         & -                        & 2.5         & -                         \\
            \hlineB{2.5}
        \end{tabular}
    }
    \caption{The number of elements and the relative error for each solver and mesh employed.}\label{tab:mesh_convergence}
\end{table*}


\section{\label{sec:Conclusions} Conclusions}
This study has highlighted the significant influence of simulation conditions on the fidelity and efficiency of modeling intraventricular cardiac flows. We have discussed that the choice of geometry, whether idealized or patient-specific, has a profound impact on both the computational complexity and the physiological accuracy of the simulations. While patient-specific models could provide detailed insights, they require higher computational resources and sophisticated data interpolation methods. Conversely, idealized geometries, although less detailed, provide sufficient accuracy to capture critical flow features such as the vortex ring at reduced computational effort.
\\
Our investigations into the flow conditions confirmed that the assumption of laminar flow is sufficient to simulate cardiac dynamics at the scale of the LV model, thus avoiding the additional complexity of turbulence models. The boundary conditions, in particular the dynamic modeling of wall motion and the careful choice of the inflow and outflow conditions, were crucial for accurately capturing the essential features of cardiac flow. The implementation of a moving wall boundary, as opposed to a static wall, proved to be an effective method for enhancing the capability of our study to reproduce key physiological behaviors, including vortex ring formation and flow recirculation.
\\
Furthermore, our results underscore the need to employ a refined mesh near the walls to resolve small-scale phenomena and dynamically adapt to wall motion, thereby improving the overall quality of the simulation.
\\
Validation of our simulations through comparisons with commercial solvers Ansys Fluent and Star-CCM+ confirmed the reliability and robustness of our approach. Both solver implementations showed comparable patterns in total kinetic energy and vortex ring formation and evolution, closely matching each other and the existing literature. This double validation reinforces the accuracy of our methodology and its applicability in realistic cardiac flow simulations.
\\
In conclusion, this study provides a comprehensive assessment of the critical conditions needed to accurately and efficiently simulate the complex dynamics within the LV model, providing insights that could be instrumental in advancing the field of cardiovascular research and the development of diagnostic and therapeutic strategies.

\section{Data Availability Statement}
The data that support the findings of this study are available upon reasonable request. 

\section*{Acknowledgements}
The authors acknowledge the grants PID2023-147790OB-I00, TED2021- 129774B-C21 and PLEC2022-009235 funded by MCIN/AEI/ 10.13039/501100011033 and by the European Union “NextGenerationEU”/PRTR, and S.L.C. acknowledges the support of Comunidad de Madrid through the call Research Grants for Young Investigators from Universidad Polit{\'e}cnica de Madrid. We extend our gratitude to our partners in  Centro Nacional de Investigaciones Cardiovasculares (CNIC). The authors gratefully acknowledge the Universidad Politécnica de Madrid (www.upm.es) for providing computing resources on Magerit Supercomputer.


\end{document}